\documentclass[
nofootinbib,
superscriptaddress,
showpacs
]{revtex4}

\usepackage{graphicx}
\usepackage{epsfig}
\usepackage{latexsym}
\usepackage{amsmath}
\usepackage{amsfonts}
\usepackage{amsxtra}

\newcommand{\msbar}{\overline{\mbox{{\sc ms}}}}
\newcommand{\wmsbar}{{\rm W}\overline{\mbox{{\sc ms}}}}
\newcommand{\MOM}{\mbox{{\sc mom}}}

\newcommand{\lwrsim}{\raise0.3ex\hbox{$<$\kern-0.75em\raise-1.1ex\hbox{$\sim$}}}
\def\krto{ {\,\,\lower .8ex\hbox {$\longrightarrow \atop k \rightarrow 0$}\,\,}}

\def\bea{\begin{eqnarray} }
\def\beq{\begin{eqnarray} }

\def\eea{\end{eqnarray}}
\def\eeq{\end{eqnarray}}

\def\eq#1{Eq.~(\ref{#1})}

\begin{document} \date{\today}

\title{Testing the OPE Wilson coefficient for $A^2$ from lattice QCD with a dynamical charm}

\author{B.~Blossier}
\author{Ph.~Boucaud} 
\affiliation{Laboratoire de Physique Th\'eorique, 
Universit\'e de Paris XI; B\^atiment 210, 91405 Orsay Cedex; France}
\author{M.~Brinet}
\affiliation{LPSC, CNRS/IN2P3/UJF; 
53, avenue des Martyrs, 38026 Grenoble, France}
\author{F.~De Soto}
\affiliation{Dpto. Sistemas F\'isicos, Qu\'imicos y Naturales, 
Univ. Pablo de Olavide, 41013 Sevilla, Spain}
\author{V.~Morenas}
\affiliation{Laboratoire de Physique Corpusculaire, Universit\'e Blaise Pascal, CNRS/IN2P3 
63177 Aubi\`ere Cedex, France}
\author{O.~P\`ene}
\affiliation{Laboratoire de Physique Th\'eorique, 
Universit\'e de Paris XI; B\^atiment 210, 91405 Orsay Cedex; France}
\author{K.~Petrov}
\affiliation{Laboratoire de
l'Accélérateur Linéaire, Centre Scientifique d'Orsay; B\^atiment 200, 91898 ORSAY Cedex, France}
\author{J.~Rodr\'{\i}guez-Quintero}
\affiliation{Dpto. F\'isica Aplicada, Fac. Ciencias Experimentales; 
Universidad de Huelva, 21071 Huelva; Spain.}

\begin{abstract}

Gluon and ghost propagators data, obtained in Landau gauge from lattice simulations with two light and two heavy dynamical quark flavours ($N_f$=2+1+1), are described here with a running formula including a four-loop perturbative  expression and a nonperturbative OPE correction dominated by the local operator $A^2$. The Wilson coefficients and their variation as a function of the coupling constant are extracted from the numerical data and compared with the theoretical expressions that, after being properly renormalized, are known at ${\cal O}(\alpha^4)$. As also $\Lambda_{\msbar}$ is rather well known for $N_f$=2+1+1, this allows for a precise consistency test of the OPE approach in the joint description of different observables. 

\end{abstract}


\maketitle





\section{Introduction}

The Green functions of a general quantum field theory contain in principle all the information of the theory. Consequently, since a few decades, considerable research effort is being devoted to construct as reliable as possible estimates of these Green functions in QCD. The main motivation for this has been to get some understanding about the infrared properties of QCD (see~\cite{Boucaud:2011ug} for a recent review), but they have been also aimed at extracting low-energy parameters as $\Lambda_{\rm QCD}$ from lattice QCD computations (see, for instance,  \cite{Alles:1996ka,Boucaud:1998bq,Becirevic:1999uc,Becirevic:1999hj,vonSmekal:1997is,Boucaud:2005xn,Boucaud:2008gn}). 
In particular, gluon and ghost propagators have been recently computed, within the ETM collaboration, 
from lattice gauge fields including $N_f$=2+1+1, two light degenerate and two heavy, dynamical quark flavours (see~\cite{Baron:2010bv,Baron:2011sf} for details of the simulations set-up) with a twisted-mass fermion action~\cite{Frezzotti:2000nk,Frezzotti:2003xj}. Those propagators were first exploited for the calculation of the $\msbar$ Strong running coupling {\it via} the Taylor-scheme coupling in Landau gauge~\cite{Blossier:2011tf,Blossier:2012ef} and, more recently, 
with pure Yang-Mills (no flavours) and two-light-flavours propagators, all together 
studied in order to investigate the effects on them from dynamical quark flavours~\cite{Ayala:2012pb}. The running coupling estimated through these propagators with 
$N_f$=2+1+1 at $\tau$ and $Z_0$ mass scales has been proven to agree pretty well with experimental 
results from $\tau$ decays~\cite{Bethke:2011tr} and "{\it world average}" from PDG~\cite{Nakamura:2010zzi}. 
In the analysis of refs.~\cite{Blossier:2011tf,Blossier:2012ef}, the invocation of nonperturbative OPE corrections is unequivocally needed to account properly for the lattice data of the coupling in the Taylor scheme. 

The OPE approach allows for the expansion of the matrix element of any non-local operator in terms of local operators, that can be conveniently organized in a hierarchy by their momentum dimensions. Then, the terms of that OPE expansion for any Green function provide with nonperturbative corrections that, after the sum rules factorization~\cite{Shifman:1978bx,Shifman:1978by}, appear coded as a coefficient to be computed in perturbation (Wilson coefficient) and the nonperturbative condensate of a local operator.  
Furthermore, a consistent OPE picture implies that, with the proper renormalization of the local operators, their condensates should be "universal" and take the same value in the OPE expansion for any different Green function.
In refs.~\cite{Blossier:2011tf,Blossier:2012ef} (also in refs.~\cite{Boucaud:2008gn,Blossier:2010ky} for $N_f$=0 and $N_f=$2 quark flavours), the Taylor-scheme running coupling is confronted to its OPE prediction to leave us with a best fit of both $\Lambda_{\rm QCD}$ and the dominant nonperturbative condensate, {\it i.e.} that of the local operator $A^2$ which has been elsewhere very much studied ({\it e.g.}, see \cite{Boucaud:2000nd,Gubarev:2000nz,Kondo:2001nq,Verschelde:2001ia,Dudal:2002pq,RuizArriola:2004en,Vercauteren:2011ze}).   It is worth to emphasize that, in constructing the Taylor-scheme coupling, the bare lattice propagators appear combined such that the cut-off dependence must disappear and any correction depending on the lattice spacing hence will vanish when approaching the continuum limit~\cite{vonSmekal:1997is}. This makes easier the confrontation of lattice data and continuum predictions but prevents the check for the universality of condensates, as gluon and ghost propagators are not to be independently analyzed.  On the other hand, 
concerning the universality of condensates, previous studies exploiting quenched  ($N_f$=0)~\cite{Boucaud:2000nd,Boucaud:2001st,DeSoto:2001qx,Boucaud:2005xn,Boucaud:2005rm,Boucaud:2008gn} 
lattice gauge configurations have shown a fair consistency for the values of this dimension-two gluon condensate extracted from the dominant nonperturbative corrections for gluon, ghost and quark propagators and the three-gluon vertex, in Landau gauge. The same have been found when comparing the condensate extracted from lattice quark propagator and the Taylor coupling with $N_f$=2 dynamical flavours~\cite{Blossier:2010ky,Blossier:2010vt}. Still, despite that consistency and the pretty convincing estimate of $\Lambda_{\rm QCD}$ obtained with the good agreement of the OPE formula and the lattice results for the Taylor coupling with $N_f=$2+1+1, the nature itself of the unavoidable nonperturbative corrections and the involved condensates is a controversial matter (there was, for instance, some recent controversial work about the notion of condensates that can be followed in refs.~\cite{Chang:2011mu,Brodsky:2012ku,Lee:2012ga} and references therein). 

In the present note, we will make one more step with respect to refs.~\cite{Blossier:2012ef,Blossier:2011tf} and will face any controversy about the nonperturbative OPE approach only by following that "{\it the proof of the pudding is in the eating}". Then, our goal here will be to test again the validity of the OPE approach, now describing gluon and ghost propagators computed in Landau gauge from realistic lattice simulations including non-degenerated strange and charm quarks ($N_f$=2+1+1). The running with momenta for both propagators is well known at the four-loop level from perturbation theory. As one deals with a very realistic simulation of QCD at the energy scale for the $\tau$ physics, $\Lambda_{\rm QCD}$ will be taken as an input from experiments with $\tau$ decays. Therefore, the deviation with respect to the perturbative running for the lattice propagators will be accurately isolated and identified to the leading OPE power correction. This will allow for a very precise comparison that will be sensitive to the running with momenta of the Wilson coefficient for the local operator $A^2$, which is known at ${\cal O}(\alpha^4)$ order. 
Thus, the deviation from the perturbation theory for gluon and ghost lattice propagators 
will be accurately accommodated within a nonperturbative OPE correction, including the running of the Wilson coefficient, by fitting a value of the condensate for each propagator and verifying that 
they both agree. {\it This is a double and very demanding check}. 

The paper is organized as follows: the nonperturbative running formulae for the propagators, after  properly defining the renormalization prescription, are derived in section \ref{sec:OPE}; the results are compared with their lattice estimates in section \ref{sec:ghgl}; the Taylor coupling is again obtained from bare gluon and ghost propagators and, for the sake of consistency, compared with the nonperturbative prediction in section \ref{sec:Taylor}; and the conclusions are finally presented in section \ref{sec:conclu}

\section{The Wilson coefficients at ${\cal O}\left(\alpha^ 4\right)$}
\label{sec:OPE}

The purpose of this section is to derive the running of the propagators including an OPE contribution dominated by the local dimension-two operator $A^2$. In particular, we focus on the precise definition of the renormalization prescription and in obtaining the Wilson coefficients at ${\cal O}(\alpha^4)$ order. We will first compute the OPE-based equations for both gluon and ghost propagators and then combine them to derive the expression describing the nonperturbative running for the Taylor coupling. 

\subsection{Two-point functions}

For the sake of generality, let us invoke a bare scalar form factor, $\Gamma(p^2,\Lambda)$, defined from any gauge-dependent two-point Green function, where $\Lambda$ is a regularization parameter and $p^2$ the momentum scale. We can formally write the following OPE expansion:
\beq\label{eq:Gammab}
\Gamma(p^2,\Lambda) \ = \ \Gamma^{\rm pert}\left(p^2,\Lambda\right) \ 
\left( 1 + \frac{\Gamma^{A^2}\left(p^2,\Lambda\right)}{\Gamma^{\rm pert}\left(p^2,\Lambda\right)} \frac{\langle A^2 \rangle_\Lambda}{4\left(N^2_c-1\right) p^2} 
\ + \ \cdots \ \right) \ ,
\eeq
which is taken to be dominated in Landau gauge by the local operator $A^2$ and where the purely perturbative contribution\footnote{Using the path integral jargon, it corresponds with the contribution to the Green function from field configurations with a null vacuum expectation value of the local operators in the OPE expansion.} is explicitly factorized. The bracket in r.h.s. of \eq{eq:Gammab} must be finite for a renormalizable theory, as the same multiplicative renormalization constant is required to remove on a same footing all the the terms in the OPE expansion. Furthermore, one can apply the dimensional regularization procedure (this implies $\Lambda=2/(4-d)=\varepsilon^{-1}$) and then, for dimensional reasons, the term proportional to $1/p^2$ cannot produce an additional pole at $d=4$ in a renormalizable theory. Therefore, the poles for the obviously divergent quantity inside the bracket 
\beq\label{eq:ratio}
\frac{\Gamma^{A^2}\left(p^2,\varepsilon^{-1}\right)}{\Gamma^{\rm pert}\left(p^2,\varepsilon^{-1}\right)}
\eeq
and those coming from the condensate $\langle A^2 \rangle_{\varepsilon^{-1}}$ kill each other. In other words, when dealing with the UV divergencies, one can renormalize the ratio of \eq{eq:ratio} as follows  
\beq\label{eq:ren}
Z_{A^2}(\mu^ 2,\varepsilon^{-1}) \frac{\Gamma^{A^2}\left(p^2,\varepsilon^{-1}\right)}{\Gamma^{\rm pert}\left(p^2,\varepsilon^{-1}\right)} \ \equiv \ \frac{c_2^{\wmsbar}\left(\frac{p^2}{\mu^2},\alpha(\mu^2)\right)}{c_0^{\MOM}\left(\frac{p^2}{\mu^2},\alpha(\mu^2)\right)} \ ,
\eeq
where the dimension-two local operator is renormalized by $Z_{A^2}$ such that 
\beq
\langle A^2 \rangle_{{\rm R},\mu^ 2} \ = \ Z_{A^2}^{-1}(\mu^2,\varepsilon^{-1}) \ \langle A^2 \rangle_{\varepsilon^{-1}} \ 
\eeq
and we prescribe the standard $\msbar$ renormalization recipe that only $1/\varepsilon$-poles and accompanying factors $\log(4\pi)$ are to be dropped away. \eq{eq:ren}'s r.h.s. stands for the definition of $c_2^{\wmsbar}$, in the "Wilson $\msbar$" scheme\footnote{This scheme is defined for the Wilson coefficient in the OPE expansion by imposing that the local operator is renormalized in $\msbar$, while the expanded Green function is in $\MOM$ scheme.} ($\msbar$)~\cite{Blossier:2010vt}, after the introduction of
\beq\label{eq:c0MOM}
c_0^{\MOM}(p^2,\mu^2) \ = \ \frac{\Gamma^{\rm pert}\left(p^2,\varepsilon^{-1}\right)}{\Gamma^{\rm pert}\left(\mu^2,\varepsilon^{-1}\right)} \ ,
\eeq
which allows to recast \eq{eq:Gammab} that now will read
\beq
\label{eq:GammaR1}
\Gamma(p^2,\varepsilon^{-1}) \ = \ \Gamma^{\rm pert}\left(p^2,\varepsilon^{-1}\right) \ 
\left( 1 +  \frac{c_2^{\wmsbar}\left(\frac{p^2}{\mu^2},\alpha(\mu^2)\right)}
{c_0^{\MOM}\left(\frac{p^2}{\mu^2},\alpha(\mu^2)\right)}
 \ \frac{\langle A^2 \rangle_{{\rm R},\mu^2}}{4\left(N^2_c-1\right) p^2} 
\ + \ \cdots \ \right) \ .
\eeq 
Thus, one can immediately notice that the renormalization of the Wilson coefficient inside the bracket and that of the two-point Green function outside are independent processes, irrespective with each other. In particular, one can also prescribe the standard $\msbar$ renormalization scheme for the two-point Green function and will obtain the following helpful relation:
\beq\label{eq:c2s}
c_2^{\msbar}\left(\frac{p^2}{\mu^2},\alpha(\mu^2)\right) \equiv
c_0^{\msbar}\left(\frac{p^2}{\mu^2},\alpha(\mu^2)\right)
\frac{c_2^{\wmsbar}\left(\frac{p^2}{\mu^2},\alpha(\mu^2)\right)}
{c_0^{\MOM}\left(\frac{p^2}{\mu^2},\alpha(\mu^2)\right)} \ ;
\eeq
that will allow for a direct link of $c_2^{\wmsbar}$ to the standard $\msbar$ Wilson coefficient computed at ${\cal O}\left(\alpha^4\right)$-order in \cite{Chetyrkin:2009kh}. 
For the purposes we pursue, the running with momentum, $p$, for the OPE nonperturbative correction inside the bracket of \eq{eq:GammaR1} can be derived by solving the following renormalization group (RG) equation
\beq\label{eq:RGE}
\left\{ -\gamma_{A^2}^{\msbar}\left(\alpha(\mu^2)\right) + \frac{\partial}{\partial\log{\mu^2}} + \beta\left(\alpha(\mu^2)\right) \frac{\partial}{\partial\alpha}  
\right\} \ 
\frac{c_2^{\wmsbar}\left(\frac{p^2}{\mu^2},\alpha(\mu^2)\right)}
{c_0^{\MOM}\left(\frac{p^2}{\mu^2},\alpha(\mu^2)\right)}
\ = \ 0 \ ,
\eeq
with
\beq\label{eq:gammaA2}
\gamma_{A^2}^{\msbar}\left(\alpha(\mu^2)\right) \ = \ \frac{d}{d\log{\mu^2}} \log{Z_{A^2}} 
\ = \ - \sum_{i=0} \gamma_i^{A^2} \left(\frac{\alpha(\mu^2)}{4\pi}\right)^{i+1} \ , 
\eeq
the anomalous dimension for the dimension-two local operator $A^2$ that can be found at 
the four-loop order in ref.~\cite{Gracey:2002yt}. This RG equation is obtained just by 
applying the logarithm derivative on the renormalization momentum, $\mu$, 
to the two hand sides of \eq{eq:ren}. The boundary condition for the solution, 
\beq\label{eq:bc}
 \frac{c_2^{\wmsbar}\left(1,\alpha(p^2)\right)}
{c_0^{\MOM}\left(1,\alpha(p^2)\right)} \ = \ 
\frac{c_2^{\msbar}\left(1,\alpha(p^2)\right)}
{c_0^{\msbar}\left(1,\alpha(p^2)\right)} \ ,
\eeq
is provided by \eq{eq:c2s} and can be borrowed from refs.~\cite{Chetyrkin:2009kh}, 
where $c_2^{\msbar}(1,\alpha(q^2))$ is given by taking $q^2=\mu^2$ in Eq.(7)/Eq.(9) 
for the gluon/ghost case, and \cite{Chetyrkin:2000dq}, where $c_0^{\msbar}(1,\alpha)$ 
for both ghost and gluon two-point Green functions is given in appendix C. 

Finally, we only need to invoke multiplicative renormalizability to extend the validity of \eq{eq:GammaR1} to two-point Green functions obtained by applying any other regularization scheme. In particular, for 
lattice regularization, we will have
\beq
\label{eq:GammaRL}
\Gamma(p^2,a^{-1}) \ = \ 
z\left(\mu^2,a^{-1}\right) \ c_0^{\MOM}\left(\frac{p^2}{\mu^2},\alpha(\mu^2)\right)
\left( 1 +  \frac{c_2^{\wmsbar}\left(\frac{p^2}{\mu^2},\alpha(\mu^2)\right)}
{c_0^{\MOM}\left(\frac{p^2}{\mu^2},\alpha(\mu^2)\right)}
 \ \frac{\langle A^2 \rangle_{{\rm R},\mu^2}}{4\left(N^2_c-1\right) p^2} 
\ + \ \cdots \ \right) \ ,
\eeq 
where $a$ is the lattice spacing and $z(\mu^2,a^{-1})$ is an overall factor that, according to \eq{eq:GammaR1}, coincide with $\Gamma^{\rm pert}(\mu^2,a^{-1})$. In order to solve \eq{eq:RGE}, one can write 
\beq\label{eq:ans}
\frac{c_2^{\wmsbar}\left(\frac{p^2}{\mu^2},\alpha(\mu^2)\right)}
{c_0^{\MOM}\left(\frac{p^2}{\mu^2},\alpha(\mu^2)\right)} \ = \
\frac{c_2^{\msbar}\left(1,\alpha(p^2)\right)}
{c_0^{\msbar}\left(1,\alpha(p^2)\right)}  
\ \zeta\left(\alpha(\mu^2),\alpha(p^2)\right) \ 
\eeq 
and solve for $\zeta(\alpha(\mu^2),\alpha(p^2))$ the same \eq{eq:RGE} with the boundary condition 
$\zeta(\alpha(p^2),\alpha(p^2))=1$. Thus, we will obtain at the order ${\cal O}(\alpha^4)$
\beq\label{eq:R}
\frac{c_2^{\wmsbar}\left(\frac{p^2}{\mu^2},\alpha(\mu^2)\right)}
{c_0^{\MOM}\left(\frac{p^2}{\mu^2},\alpha(\mu^2)\right)}
 & = &
\left( \frac{\alpha(p^2)}{\alpha(\mu^2)}\right)^{\frac{27}{100}} \ 12 \pi \alpha(\mu^2) 
\left(1 - 0.54993 \ \alpha(\mu^2) - 0.14352 \ \alpha^2(\mu^2) - 0.07339 \ \alpha^3(\mu^2) \right) 
\nonumber \\
&\times & 
\left\{ \begin{array}{lr}
\left(1 + 1.31808 \ \alpha(p^2) +  1.51845 \ \alpha^2(p^2) + 3.04851 \ \alpha^3(p^2) \right) & \ \ \Gamma \equiv G \\
\left(1 +   0.89698 \ \alpha(p^2) +  0.84882 \ \alpha^2(p^2) +  0.87056 \ \alpha^3(p^2) \right) & \ \ \Gamma \equiv F 
\end{array}
\right. \ ,
\eeq
where $G$ and $F$ stand, respectively, for the gluon and ghost dressing functions and we take $\alpha \equiv \alpha_T$, the Taylor coupling, for the expansion. 

\subsection{The Taylor coupling}

As the Taylor coupling is directly obtained from the lattice in terms of gluon and ghost dressing functions, up to lattice artefacts corrections, it follows~\cite{Boucaud:2008gn},
\beq\label{eq:alphaTdef}
\alpha_T(p^2) \ = \ \frac{g_0^2(a^{-1})}{4\pi} \ F^2(p^2,a^{-1}) \ G(p^2,a^{-1})  \ .
\eeq
Then, one can combine eqs.~(\ref{eq:GammaRL},\ref{eq:R}) and be left with
\beq\label{eq:alphaTNP}
\alpha_T(p^2) \ = \ \alpha_T^{\rm pert}(p^2) \left( 1 +  9 \ 
R\left(\alpha_T^{\rm pert}(p^2),\alpha_T^{\rm pert}(\mu^2)\right) 
\left(\frac{\alpha(p^2)}{\alpha(\mu^2)}\right)^{\frac{27}{100}} \ \frac{g^2(\mu^2) \langle A^2 \rangle_{{\rm R},\mu^2}}{4\left(N^2_c-1\right) p^2} \ + \ \cdots \right)
\eeq
where
\beq\label{eq:Ra0}
R\left(\alpha,\alpha_0\right) \ = \ \left( 1 + 1.03735 \ \alpha + 1.07203 \ \alpha^2 + 1.59654 \ \alpha^3 \right) \left(1 - 0.54993 \ \alpha_0 - 0.14352 \ \alpha_0^2 -  0.07339 \ \alpha_0^3 \right) \ .
\eeq
In obtaining \eq{eq:alphaTNP}, the purely perturbative contribution, $\alpha_T^ {\rm pert}$, appears factored out of the bracket because of the following result,
\beq\label{eq:relMOM}
\left( F_0^{\rm MOM}\left(\frac{p^2}{\mu^ 2},\alpha(\mu^ 2)\right) \right)^2 \ 
 G_0^{\rm MOM}\left(\frac{p^2}{\mu^ 2},\alpha(\mu^ 2)\right) \ = \ 
\frac{\alpha_T^{\rm pert}(p^2)}{\alpha_T^{\rm pert}(\mu^2)} \ , 
\eeq
where $c_0^{\rm MOM}$, defined in \eq{eq:c0MOM}, is here dubbed $G_0^{\rm MOM}$ for the gluon case ($\Gamma \equiv G$) and $F_0^{\rm MOM}$ for the ghost ($\Gamma \equiv F$). 
\eq{eq:relMOM} can be straightforwardly derived from the nonperturbative definition of the Taylor coupling by \eq{eq:alphaTdef} but also proved at any order in perturbation. For the latter, one only needs the 
gluon and ghost anomalous dimension in MOM scheme and beta function in Taylor scheme to compute 
in perturbation $F_0^{\rm MOM}$, $G_0^{\rm MOM}$ and $\alpha_T^{\rm pert}$ and then verify  
\eq{eq:relMOM} order by order. 

The reader might have noted that coefficients in \eq{eq:Ra0} appear to be slightly different with respect to those in Eq.~(6) of ref.~\cite{Blossier:2011tf}. This discrepancy is explained by the different boundary conditions used to solve \eq{eq:RGE} in this paper, \eq{eq:bc}, and in ref.~\cite{Blossier:2011tf}, where the denominator of \eq{eq:bc}'s r.h.s. is approximated by 1. 
As Eq.~(6) of ref.~\cite{Blossier:2011tf} is applied to estimate the strong coupling in ref.~\cite{Blossier:2012ef}, replacing Eq.~(6) of~\cite{Blossier:2011tf} by \eq{eq:Ra0} would affect the strong coupling estimates, although with an almost negligible impact: the systematic deviation for $\Lambda_{\msbar}^{N_f=4}$ would increase it roughly by 1 \%, {\it i.e.} 328(18) MeV instead of 324(17) MeV.  

\section{Gluon and ghost propagators}
\label{sec:ghgl}

The analysis to be done now consists in applying \eq{eq:GammaRL} with (\ref{eq:ans},\ref{eq:R}) to 
describe the lattice gluon and ghost propagators that have been computed for the running coupling study of 
ref.~\cite{Blossier:2012ef} and used for the study of dynamical quark flavour effects in ref.~\cite{Ayala:2012pb}, 
\beq
 \Delta^{ab}_{\mu\nu}(q)&=&\left\langle A_{\mu}^{a}(q)A_{\nu}^{b}(-q)\right\rangle
 = \delta^{ab}\left(\delta_{\mu\nu}-\frac{q_\mu q_\nu}{q^2}\right) \frac{G(q^2)}{q^ 2} \;,
\nonumber \\
F^{ab}(q^2) &=& \frac 1 V \ \left\langle \sum_{x,y}
 \exp[iq\cdot(x-y)] \left( M^{-1} \right)^{ab}_{xy} \right\rangle
 =\delta^{ab}\frac{F(q^2)}{q^2} \;;
\label{green}
\eeq
where $A_\mu^ a$ is the gauge field and $M^{ab}\equiv \partial_\mu D^{ab}_\mu$ is the Fadeev-Popov operator. The reader interested in details about the computation are referred to section II and III of ref.~\cite{Ayala:2012pb}, while here we will just analyse the results for gluon and ghost dressing functions, defined by \eq{green} and obtained from the lattice simulations in \cite{Blossier:2012ef} and described in Tab.~\ref{tab:setup}, after the $O(4)$-breaking lattice artefacts have been cured by the so-called $H(4)$-extrapolation procedure~\cite{Becirevic:1999uc,deSoto:2007ht}. These estimates for bare gluon and ghost dressing functions appear plotted in Fig.~\ref{fig:green}. 

\begin{table}[ht]
\begin{tabular}{|c|c|c|c|c|c|c|}
\hline
$\beta$ & $\kappa_{\rm crit}$ & $a \mu_l$ & $a \mu_\sigma$ & $a \mu_\delta$ &
$(L/a)^3\times T/a$ & confs. \\
\hline
1.90 & 0.1632700 & 0.0040 & 0.150 & 0.1900 & $32^3\times 64$ & 50 \\	
1.95 & 0.1612400 & 0.0035 & 0.135 & 0.1700 & $32^3\times 64$ & 50 \\
2.10 &  0.1563570 & 0.0020 & 0.120 & 0.1385 & $48^3\times 96$ & 100 \\  
\hline
\end{tabular}
\caption{Lattice set-up parameters for the ensembles we used in this paper: $\kappa_{\rm crit}$ is the critical value for the standard hopping parameter for the bare untwisted mass; $\mu_l$
stands for the twisted mass for the two degenerated light quarks,
while $\mu_\sigma$ and $\mu_\delta$ define the heavy quarks
twisted masses; the last column indicates the number of gauge
field configurations exploited. This implies that the strange quark mass is roughly set to 95 MeV and 
the charm one to 1.51 GeV (in $\msbar$ at 2 GeV), while degenerate light quark masses range from 20 to 50 MeV (The lightest pseudoscalar masses approximately range from 270 to 510 MeV).}
\label{tab:setup}
\end{table}

\subsection{Removing $O(4)$-invariant lattice artefacts}
\label{subsec:remove}

\begin{figure}[th]
 \begin{center}
	\begin{tabular}{cc}
    \includegraphics[width=8.5cm]{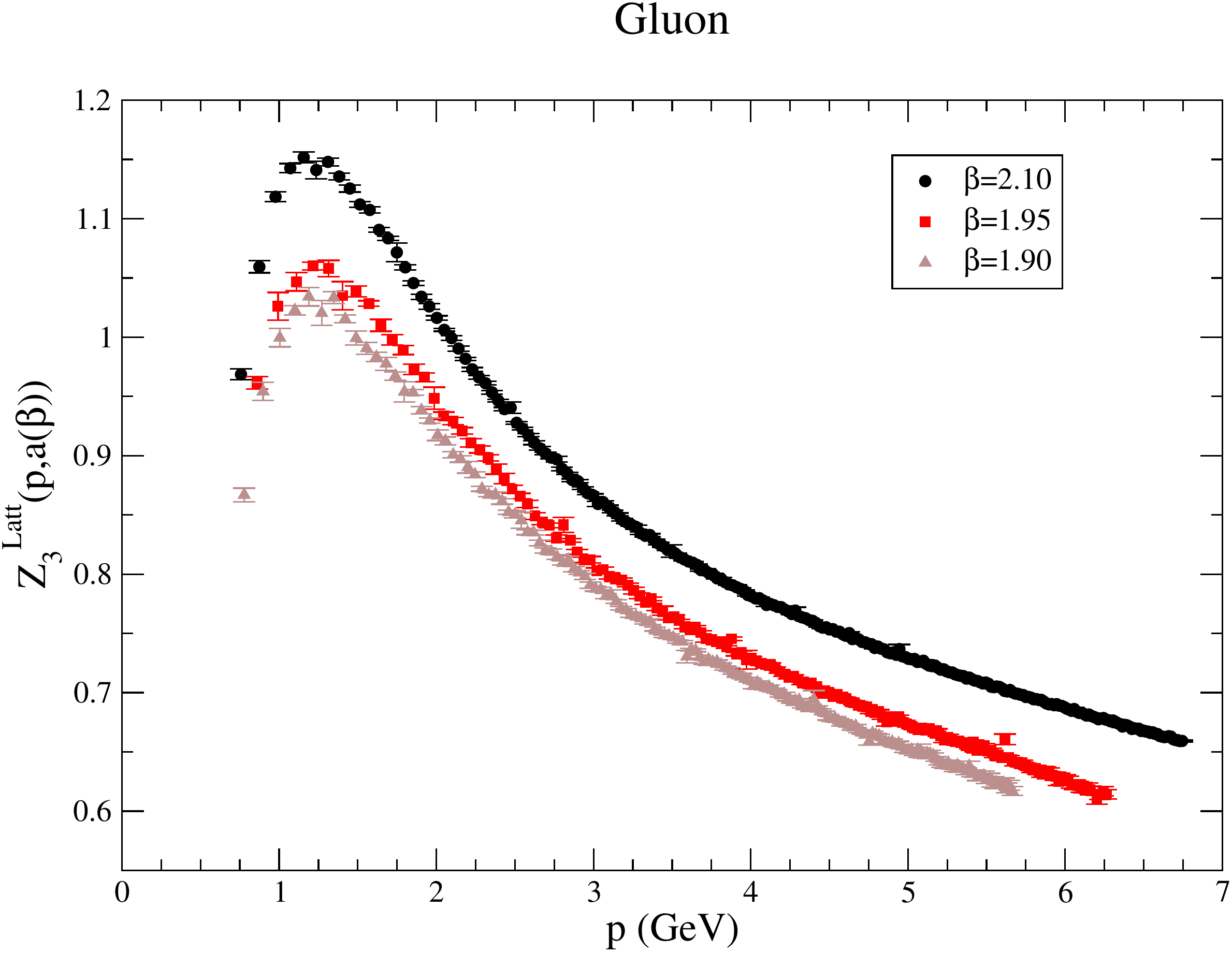} & 
    \includegraphics[width=8.5cm]{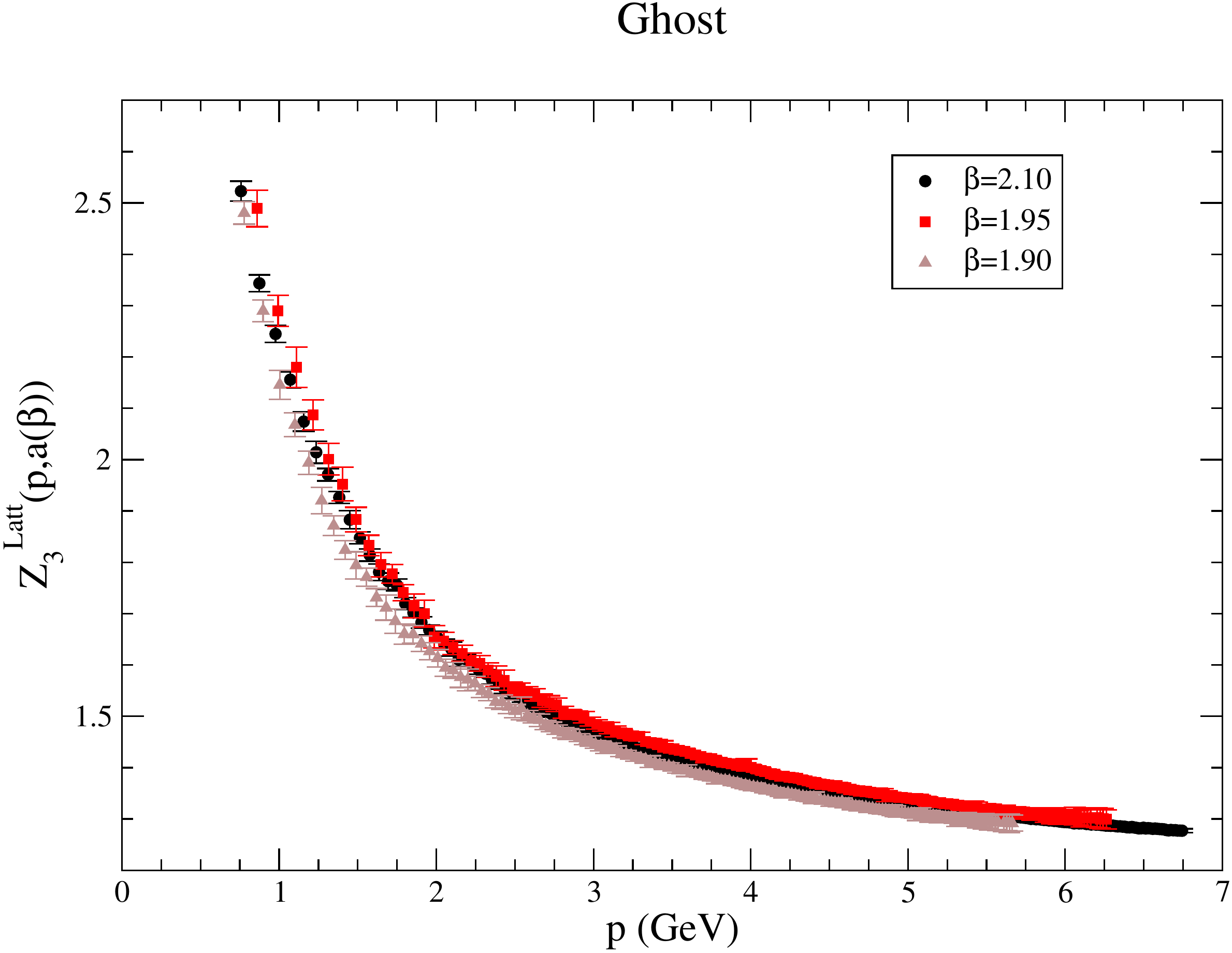} 
    \\
     \includegraphics[width=8.5cm]{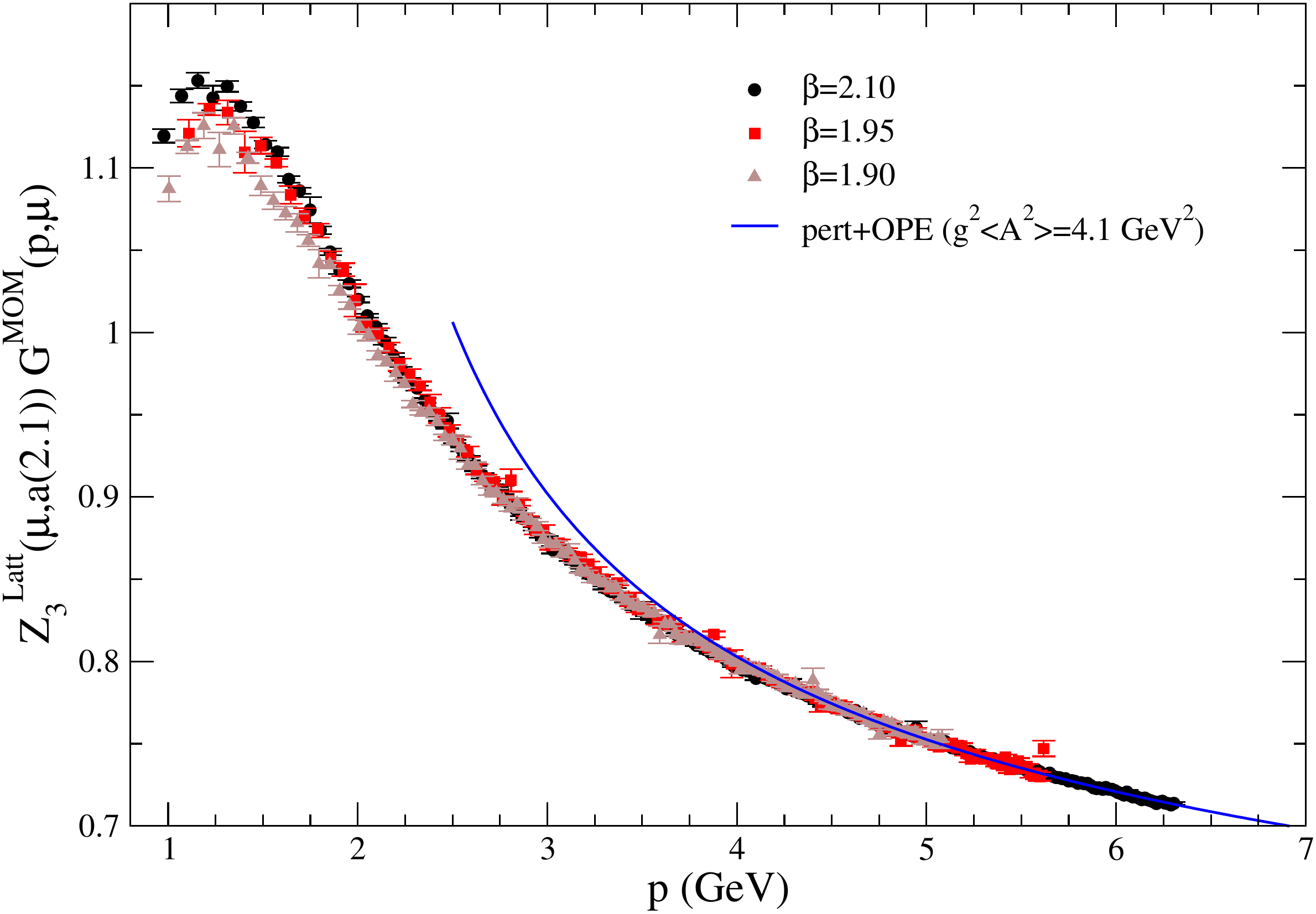} & 
    \includegraphics[width=8.5cm]{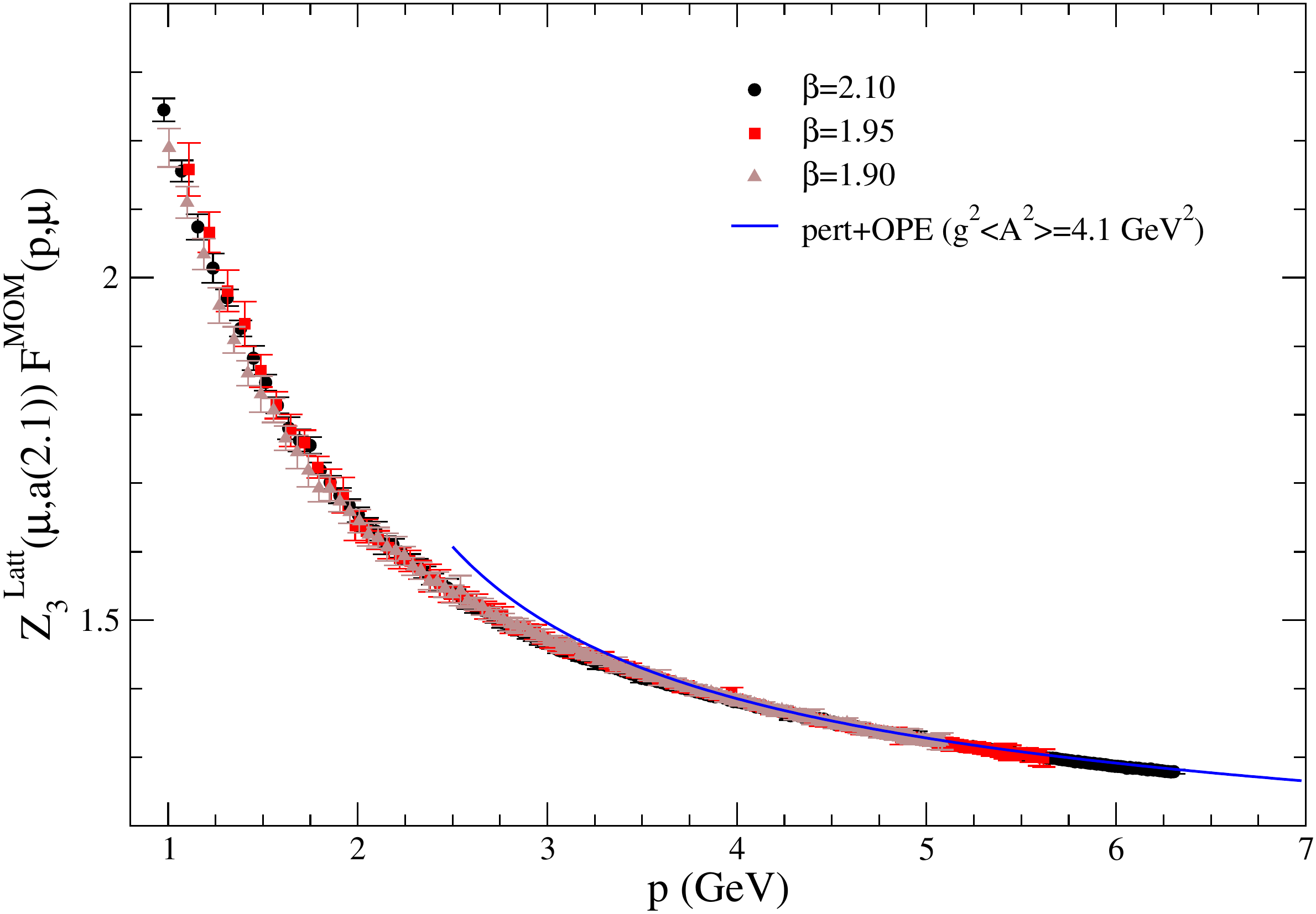} 
   
	\end{tabular}
  \end{center}
\caption{\small Bare gluon and ghost propagators as they result from lattice computations before (top) and 
after being cured of $O(4)$-invariant lattice artefacts and properly rescaled to appear superimposed (bottom). The blue solid line correspond to the best fit with the OPE formula.
}
\label{fig:green}
\end{figure}

The gluon and ghost propagator lattice data shown in the upper plots of Fig.~\ref{fig:green} are still affected by $O(4)$-invariant lattice artefacts. To treat these artefacts, we will apply the strategy that, in the following, will be described for the gluon propagator case. 

---------------------------------------

\begin{figure}[htb]
\begin{center}
\begin{tabular}{cc}
	\includegraphics[width=8.5cm]{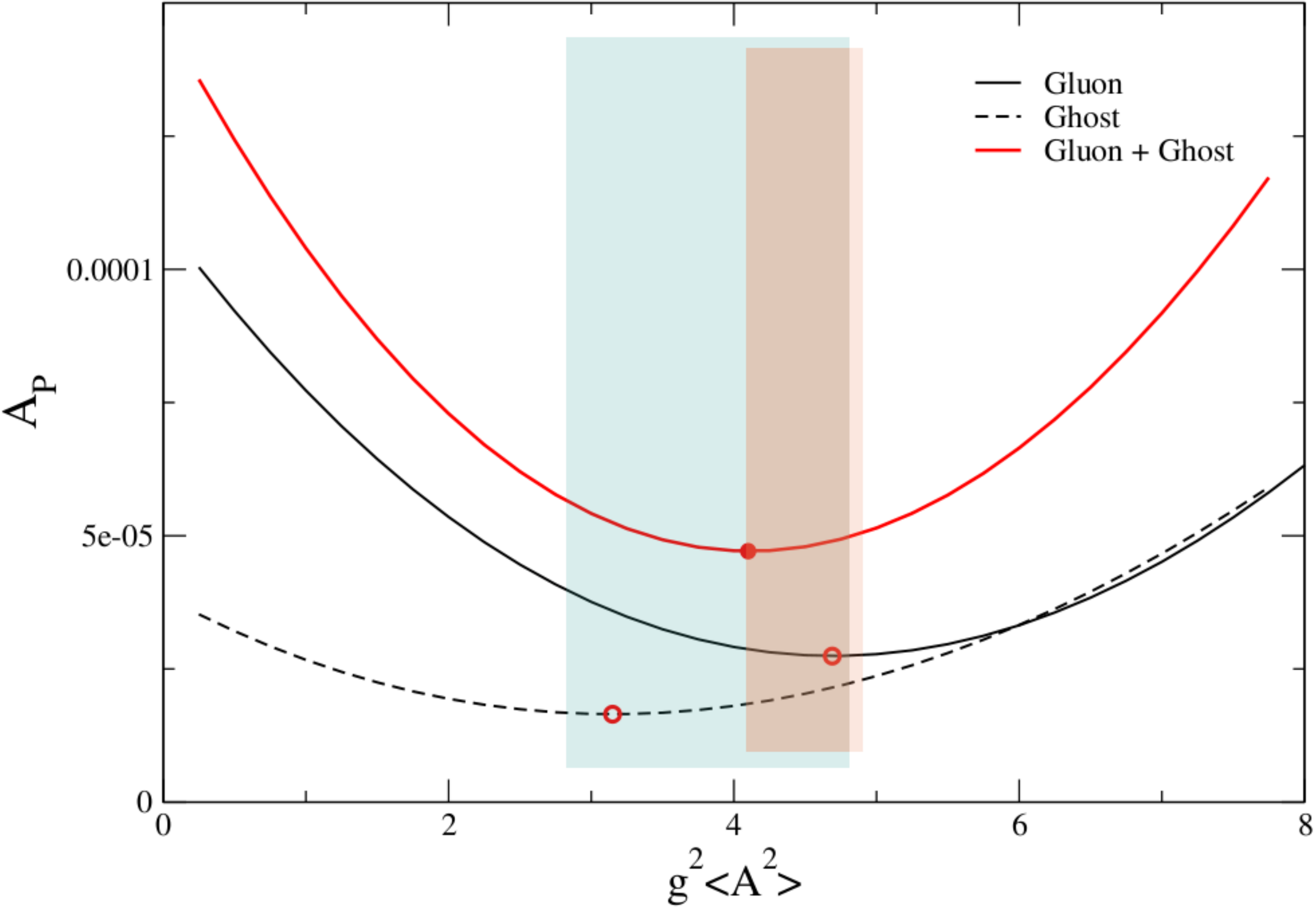}      
     & 
     \includegraphics[width=8.5cm]{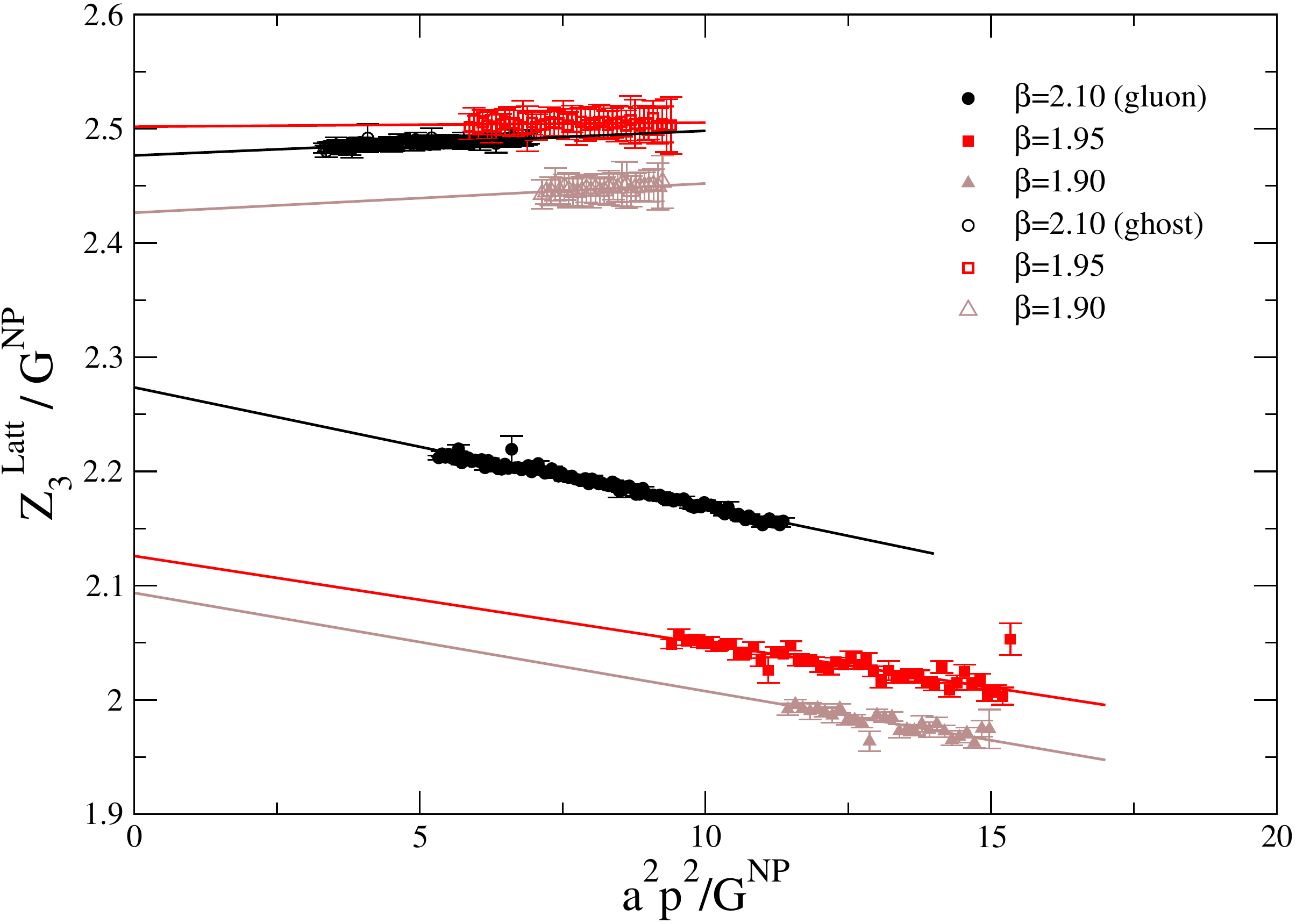}
\end{tabular}
\end{center}
\caption{\small (Left) The parameter $A_p$, related to the area between polynomial fits of the dressings for each two data sets (see appendix), in terms of the value for the gluon condensate, $g^2 \langle A^2 \rangle$. The black solid line corresponds to the gluon, the black dotted to the ghost case and the red solid one to the sum of both. The areas shadowed in blue and brown correspond to the 1-$\sigma$ allowed regions according to refs.~\cite{Blossier:2012ef} and \cite{Blossier:2011tf}, respectively. (Right) Linear fits, as suggested by \eq{eq:art}, of $Z_3^{\rm Latt}/G^{\rm NP}$ (gluon) and $\widetilde{Z}_3^{\rm Latt}/G^{\rm NP}$ (ghost) in terms of $a^2p^2/G^{\rm NP}$, where $g^2 \langle A^2 \rangle$ takes the total minimum shown in the left plot, for the three lattice data sets.
}
\label{fig:art}
\end{figure}

As discussed in ref.~\cite{Blossier:2011tf,Blossier:2012ef,Blossier:2010vt} for the cases of 
Taylor coupling and the vector quark propagator renormalization constant, one can properly parametrize the 
remaining $O(4)$-invariant lattice artefacts as follows:
\beq\label{eq:Z3g}
Z_3^{\rm Latt}(p^2,a^{-1}) \ = \ G(p^2,a^{-1}) + c_{a2p2} \ a^2 p^2 \ ,
\eeq
where $Z_3^{\rm Latt}$ is the gluon propagator renormalization constant in MOM scheme which corresponds to the lattice gluon dressing function. In r.h.s., the artefacts-free gluon dressing is noted as $G$ and, in the appropriate range of momenta, is supposed to be given by Eqs.(\ref{eq:GammaRL}-\ref{eq:R}). Then, \eq{eq:Z3g} can be recast to read
\beq\label{eq:art}
\frac{Z_3^{\rm Latt}(p^2,a^{-1})}{G^{\rm NP}\left(\frac{p^2}{\mu^ 2},\alpha(\mu^ 2)\right)} \ = \
z(\mu^2,a^{-1}) \ + \ c_{a2p2} \ \frac{a^2 p^2}{G^{\rm NP}\left(\frac{p^2}{\mu^ 2},\alpha(\mu^ 2)\right)}
\eeq
with 
\beq
\label{eq:GNP}
G^{\rm NP}\left(\frac{p^2}{\mu^ 2},\alpha(\mu^ 2)\right) \ = \ 
G_0^{\MOM}\left(\frac{p^2}{\mu^2},\alpha(\mu^2)\right)
\left( 1 +  
\frac{c_2^{\wmsbar}\left(\frac{p^2}{\mu^2},\alpha(\mu^2)\right)}
{c_0^{\MOM}\left(\frac{p^2}{\mu^2},\alpha(\mu^2)\right)}
 \ \frac{\langle A^2 \rangle_{{\rm R},\mu^2}}{4\left(N^2_c-1\right) p^2} 
\ + \ \cdots \ \right) \ ;
\eeq 
where $c_2^{\wmsbar}/c_0^{\MOM}$ is given by \eq{eq:R}'s r.h.s. for $\Gamma \equiv G$ and where $G_0^{\rm MOM}$, defined by 
\eq{eq:c0MOM}'s r.h.s. again with $\Gamma \equiv G$, can be computed 
at the four-loop order with the help of the perturbative expansions for the 
gluon propagator MOM anomalous dimension and Taylor scheme beta function~\cite{Chetyrkin:2000dq}. 
Provided that $G^{\rm NP}$ describes properly the nonperturbative running of the gluon dressing over 
a certain range of momenta, \eq{eq:art} can be used to make a linear fit of data within this range~\footnote{At lower momenta, 
the range is bounded by $p=4.5$ GeV, below which the lattice data deviate from the nonperturbative prediction only including the OPE leading power correction. At large momenta, the upper bound is provided by $a^2p^2 \sim 3.5$, above where the impact of higher-order lattice artefacts cannot be neglected any longer and results, for instance, in deviations from the linear behavior suggested by \eq{eq:art}.} for the three exploited data sets simulated at three different bare couplings, $g_0^2(a^{-1})=6/\beta$. 
This will allow for the determination of the overall factor, $z(\mu,a^{-1})$, for each data set
and the coefficient $c_{\rm a2p2}$, thought to be roughly the same~\footnote{By dimensional arguments, the coefficient $c_{\rm a2p2}$ might only depend logarithmically on the lattice spacing $a(\beta)$. Similar corrections, there inspired on the lattice perturbative expansions for the  propagators, are applied in ref.~\cite{Sternbeck:2012qs}.} for any $\beta$ (see Fig.~\ref{fig:art} and tab.~\ref{tab:res}).
Once $c_{\rm a2p2}$ known, one can remove the $O(4)$-invariant artefacts away from the lattice data for 
all available momenta, by applying \eq{eq:Z3g}, and be left with the continuum nonperturbative running for the gluon dressing; 
while the overall factors $z(\mu,a^{-1})$ can be used to rescale the data obtained at different $\beta$ such that 
they all will follow the same curve (see down plots of fig.~\ref{fig:green}).

As $G^{\rm NP}$ depends on $\Lambda_{\overline{\rm MS}}$ and the gluon condensate, 
$g^2\langle A^2 \rangle$, their values have to be known prior to the removal of $O(4)$-invariant artefacts. 
In practice, we will take $\Lambda_{\overline{\rm MS}}$ to be known from $\tau$ decays 
($\alpha_{\msbar}(m_\tau^2)=0.334(14)$~\cite{Bethke:2011tr}, {\it i.e.} $\Lambda_{\overline{\rm MS}}=311$ MeV for $N_f=4$) 
and will search for a value of $g^2\langle A^2 \rangle$ such that one gets the best matching of the rescaled gluon dressings 
from the three data sets. The criterion to define the best matching, that is explained in appendix, implies the minimization of certain parameter, $A_p$, computed with the square of 
the area between polynomial fits of the gluon dressing for each two data sets. 

\begin{table}[t]
\begin{center}
\begin{tabular}{|c|c|c|}
\hline
& gluon & ghost \\
\hline
$g^2 \langle A^2 \rangle$ [GeV$^2$] & 4.7 & 3.2 \\
\hline
\begin{tabular}{c}
\rule[0cm]{0.95cm}{0cm}
$\beta$  
\rule[0cm]{0.95cm}{0cm}
\\
\hline
$z(\mu^2,a^{-1})$ 
 \\
\hline
$c_{a2p2}$
\end{tabular}
&
\begin{tabular}{c|c|c}
 2.10 & 1.95 & 1.90 \\
\hline
 0.6495(5) & 0.6068(26) & 0.5971(30) \\
 \hline
 -0.0097(2) & -0.0071(8) & -0.0080(8)
\end{tabular}
&
\begin{tabular}{c|c|c}
 2.10 & 1.95 & 1.90 \\
\hline
 1.215(6) & 1.229(20) & 1.193(31) \\
\hline
0.0002(19) & -0.0012(44) & 0.0011(69) 
\end{tabular}
\\ 
\hline
\hline
$g^2 \langle A^2 \rangle$ [GeV$^2$] & 4.1 & 4.1 \\
\hline
\begin{tabular}{c}
$z(\mu^2,a^{-1})$ 
 \\
\hline
\rule[0cm]{0.75cm}{0cm}
$c_{a2p2}$
\rule[0cm]{0.75cm}{0cm}
\end{tabular}
&
\begin{tabular}{c|c|c}
 0.6531(5) & 0.6106(26) & 0.6014(30) \\
 \hline
 -0.0104(2) & -0.0077(8) & -0.0086(8)
\end{tabular}
&
\begin{tabular}{c|c|c}
 1.206(6) & 1.218(19) & 1.182(30) \\
\hline
0.0022(19) & 0.0004(44) & 0.003(7) 
\end{tabular}
\\ 
\hline
\end{tabular}
\end{center}
\caption{\small The overall factors, $z(\mu^2,a^{-1})$, and the coefficients $c_{a2p2}$, obtained as explained in the subsection \ref{subsec:remove}, for optimal values of $g^2\langle A^2 \rangle$. As discussed there, $c_{a2p2}$ is thought not to depend very much on $\beta$ and 
so is fairly obtained from the results shown here. The renormalization point, $\mu$, for the overall factor is fixed at 10 GeV. Statistical errors have been computed by applying the Jackknife method. The errors for the determination of the optimal values of $g^2\langle A^2 \rangle$ are also discussed in \ref{subsec:remove} and can be found in next Tab.~\ref{tab:comp}.}
\label{tab:res}
\end{table}

Thus, after applying the above described procedure to sweep the $O(4)$-invariant artefacts away from data, we obtain the minima for $A_p$ in terms of $g^2 \langle A^2 \rangle$ and the results for $c_{a2p2}$ and $z(\mu^2,a^{-1})$ shown in Tab.~\ref{tab:res}. In the left plot of Fig.~\ref{fig:art}, the minima of $A_p$ for both gluon and ghost dressing functions are shown to take place at values for $g^2 \langle A^2 \rangle$ that are fairly compatible with the estimate of \cite{Blossier:2012ef}: 3.8(1.0); and with that of \cite{Blossier:2011tf}: 4.5(4). For the sake of a proper comparison, one needs to estimate the uncertainty in determining the best value of $g^2 \langle A^2 \rangle$ by the minimization of $A_p$. To this purpose, the Jackknife procedure can be applied and gives for the gluon case the following result: $g^2\langle A^2 \rangle = 4.7(1.6)$ GeV$^2$. In the ghost case, as the coefficient $c_{a2p2}$ is found to be negligible (see Tab.~\ref{tab:res}), one can futhermore take $c_{a2p2}=0$ and so avoid very large errors due to strong correlations between this coefficient and both $z(\mu^2,a^{-1})$ and $g^2 \langle A^2 \rangle$. Thus, we will obtain $g^2\langle A^2 \rangle = 3.1(1.1)$ GeV$^2$. Those results appear collected in Tab.~\ref{tab:comp}.

\begin{table}[ht]
\begin{center}
\begin{tabular}{|c|c|c|c|c|}
\hline
& gluon & ghost & $\alpha_T$ \cite{Blossier:2011tf} & $\alpha_T$ \cite{Blossier:2012ef} 
\\
\hline 
$g^2 \langle A^2 \rangle$ [GeV$^2$] & 4.7(1.6) & 3.1(1.1) & 4.5(4) & 3.8(1.0) 
\\
\hline
\end{tabular}
\end{center}
\caption{\small Results for $g^2\langle A^2 \rangle$ independently obtained from gluon and ghost propagators in this paper and from the Taylor coupling analysis in refs.~\cite{Blossier:2011tf,Blossier:2012ef}.}
\label{tab:comp}
\end{table}

Finally, one can impose the condensates to be the same for both gluon and ghost propagators and a total minimum for the sum of gluon and ghost $A_p$'s will be found for $g^2\langle A^2 \rangle=4.1$ GeV$^2$, that appears to be again in pretty good agreement with the previous results in refs.~\cite{Blossier:2011tf,Blossier:2012ef}. In the following, we will use the artefacts-free gluon and ghost dressing functions, obtained with this last value for the gluon condensate, shown in Fig.~\ref{fig:green}.

\subsection{Testing the OPE Wilson coefficients from data}

\begin{figure}[hbt]
\begin{center}
\begin{tabular}{cc}
	\includegraphics[width=8.5cm]{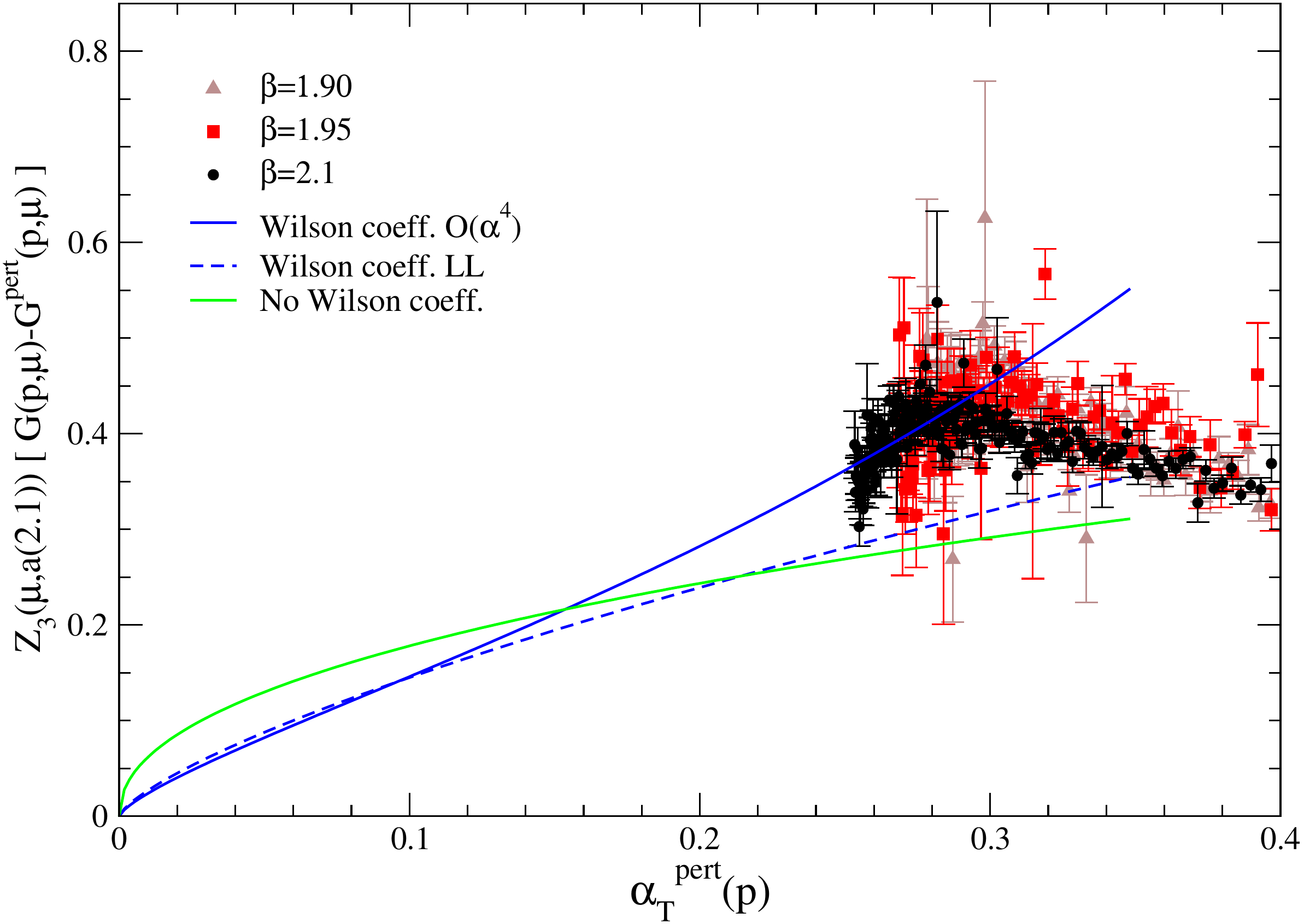}      
     & 
     \includegraphics[width=8.5cm]{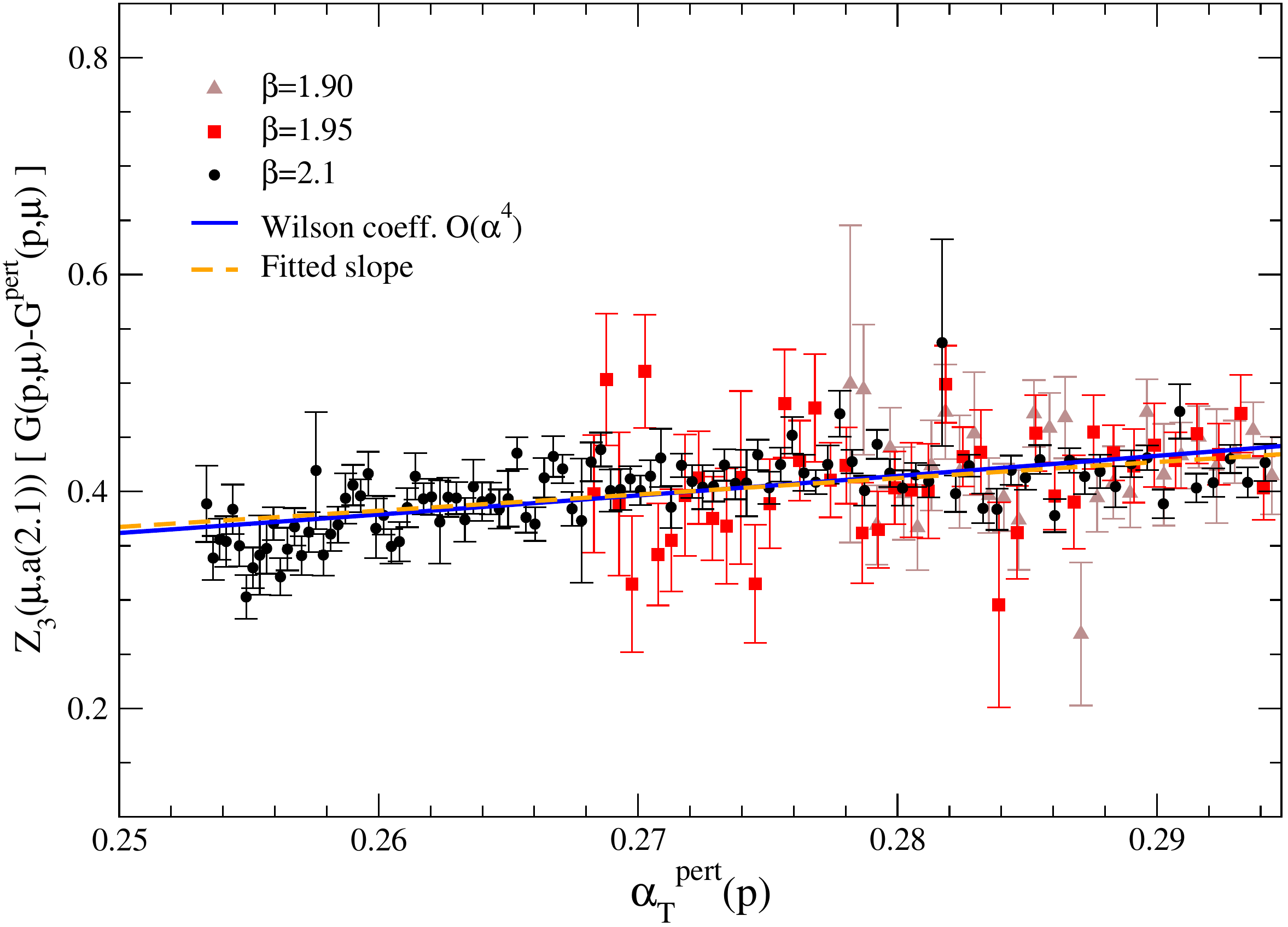} 
     \\
     	\includegraphics[width=8.5cm]{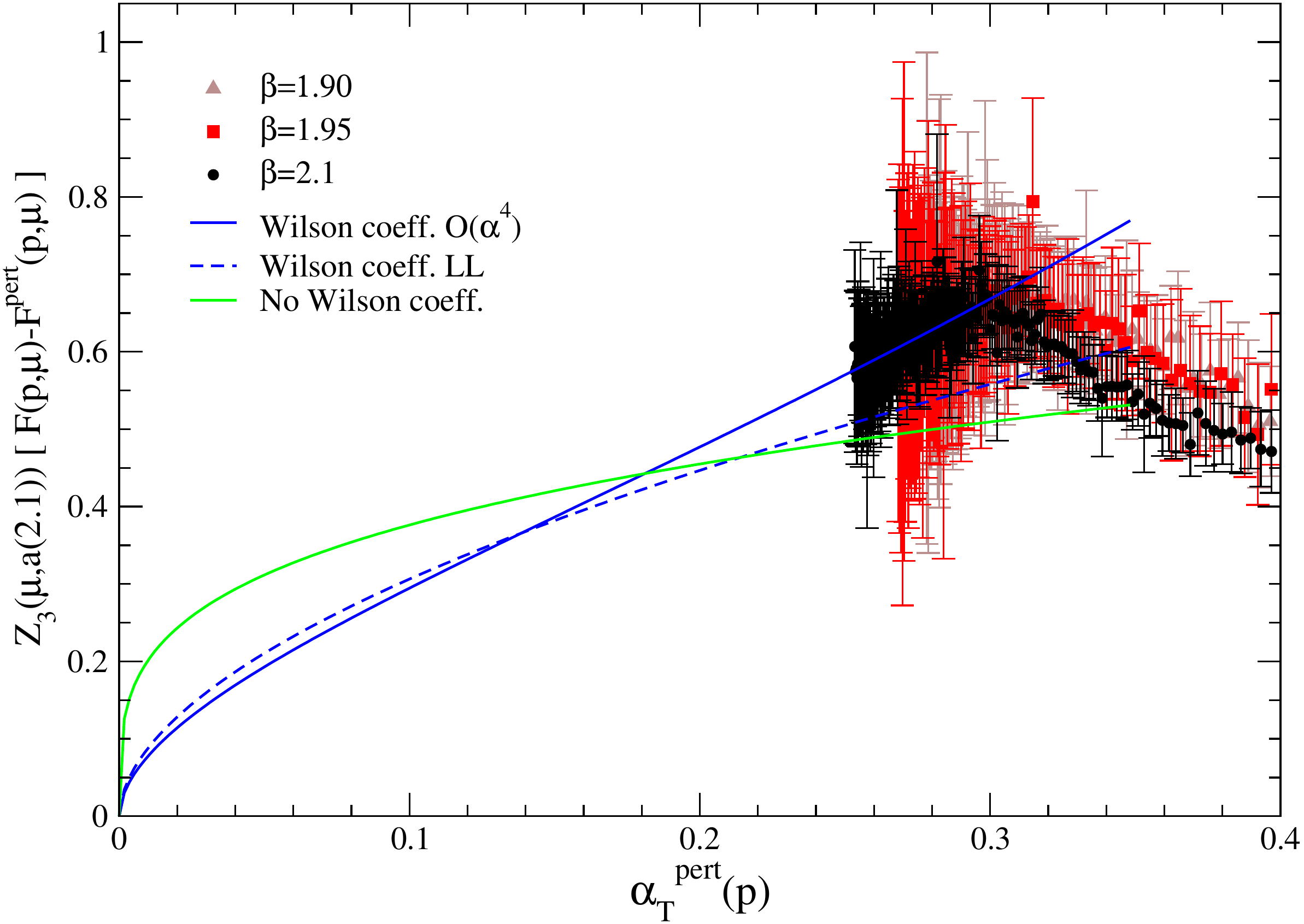}      
     & 
     \includegraphics[width=8.5cm]{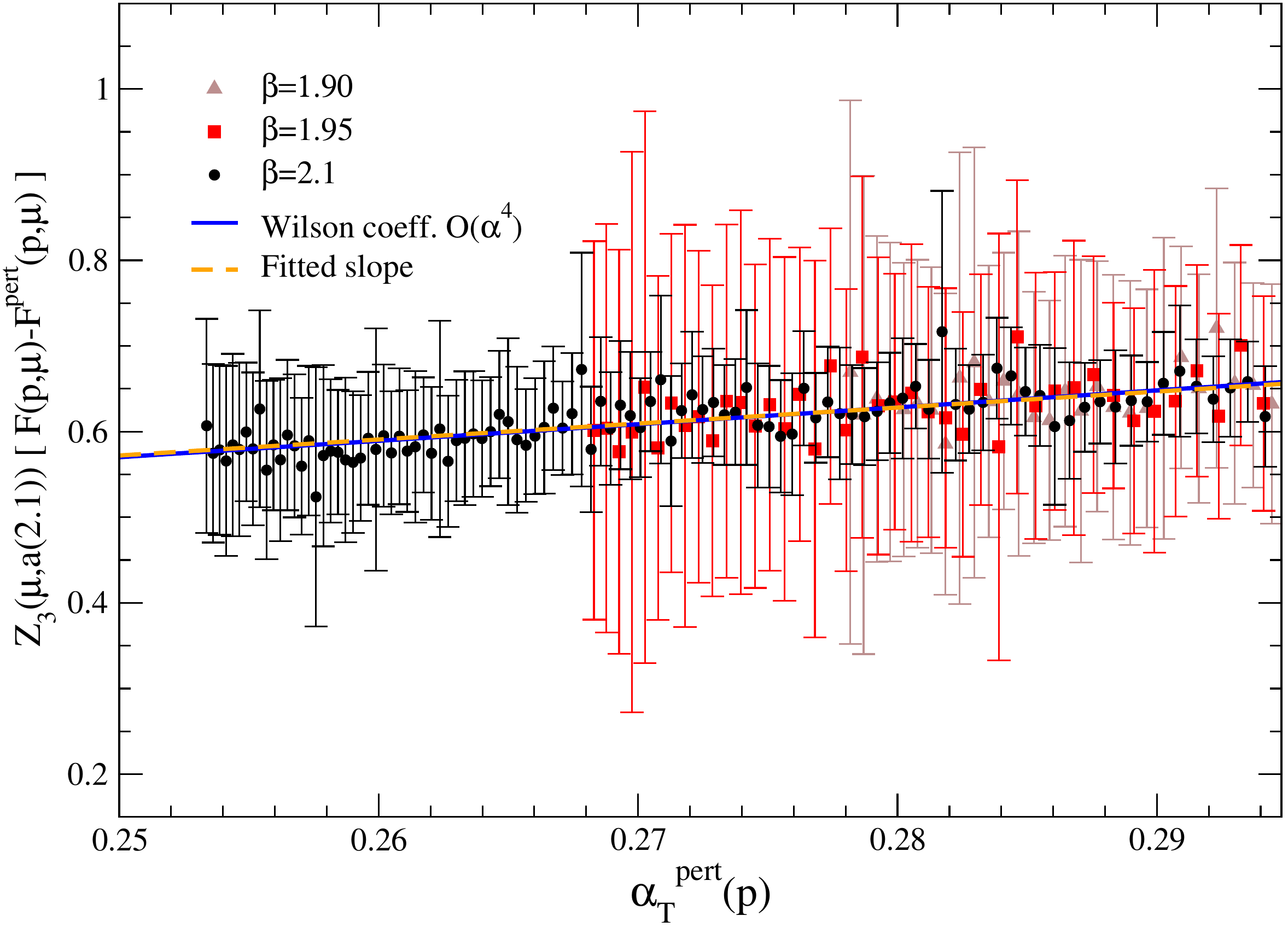} 
\end{tabular}
\end{center}
\caption{\small {\it Scrutinizing the Green functions data to isolate the running for the Wilson coefficients}. The left plots shows the difference of the rescaled lattice data for the gluon (top) and ghost (bottom) dressing functions and their perturbative estimate, 
multiplied by the square of momentum, plotted versus the perturbative Taylor coupling at the same momentum. For perturbative evaluations, $\Lambda_{\msbar}$ is taken to be known from $\tau$ decays. Blue dotted and green solid lines respectively correspond to the leading logarithm approximation for Wilson coefficient and to its suppression. 
The right plots only show the region, from 4.5 to 6.3 GeV where the behavior predicted by the OPE analysis in sec.~\ref{sec:OPE} is observed. The dotted orange line shows a linear fit of data that results to agree very well with the predicted running.
}
\label{fig:Wilson}
\end{figure}

At this point we are in a position to extract directly from the numerical data the running induced by 
Wilson coefficients, obtaining thus their behaviour as a function of the coupling constant and making  comparisons with the theoretical expectations, derived in sec.~\ref{sec:OPE}. The ability of the OPE to describe consistently and accurately different Green functions can then be tested. Such is a smoking gun for the reliability of the OPE nonperturbative approach. To this goal, \eq{eq:GNP} can be recast as follows,
\beq\label{eq:W1}
p^2 \ \left[ \frac{G(p^2,a^{-1})}{z(\mu^2,a^{-1})} - G^{\rm MOM}_0\left(\frac{p^2}{\mu^2},\alpha(\mu^2)\right) \right] \ = \
G^{\rm MOM}_0\left(\frac{p^2}{\mu^2},\alpha(\mu^2)\right) \frac{c_2^{\wmsbar}\left(\frac{p^2}{\mu^2},\alpha(\mu^2)\right)}
{c_0^{\MOM}\left(\frac{p^2}{\mu^2},\alpha(\mu^2)\right)} 
\frac{\langle A^2 \rangle_{{\rm R},\mu^2}}{4\left(N^2_c-1\right)}  \ ;
\eeq
where the l.h.s. is to be computed from the artefacts-free dressing function, $G$, defined by \eq{eq:Z3g} and obtained, as explained 
in the previous subsection, by requiring the optimal matching of data from the three lattice data sets at different $\beta$. On \eq{eq:W1}'s r.h.s., the running with momenta can be easily expressed as a perturbative series in terms of the Taylor coupling (see, for instance, \eq{eq:R}), being proportional to the gluon condensate, $g^2 \langle A^2 \rangle$, which has been also obtained in the previous section through the best-matching criterion. Then, one can plot \eq{eq:W1}'s l.h.s. in terms of $\alpha_T$ and compare with r.h.s.'s prediction. This can be seen in Fig.~\ref{fig:Wilson} for the gluon (top) and ghost (bottom) cases. In view of this, we can conclude that: (i) The nonperturbative corrections are totally unavoidable, as data for \eq{eq:W1}'s l.h.s. are clearly not compatible with zero; (ii) the Wilson coefficient derived at the order ${\cal O}(\alpha^4)$ in sec.~\ref{sec:OPE} gives a high-quality fit of data (blue solid lines in both plots); (iii) had we not included the Wilson coefficient running ({i.e.}, dropped $c_2^{\wmsbar}/c_0^{\MOM}$ away from \eq{eq:W1}'s r.h.s.), or just included its leading logarithm approximation, a flatter slope would be obtained not describing properly the data (blue dotted and green solid lines). It is worthwhile to emphasize that when applying a simple linear fit of data within the appropriate range of momenta ({\it i.e.}, $0.250 < \alpha_T < 0.295$), one obtains the orange dotted line shown in the right plots, the slope of which agrees pretty well with the local one resulting from the full r.h.s. of \eq{eq:W1}, in both gluon and ghost cases.

\section{Reconstructing the Taylor coupling}
\label{sec:Taylor}

\begin{figure}[hbt]
\begin{center}
	\includegraphics[width=12cm]{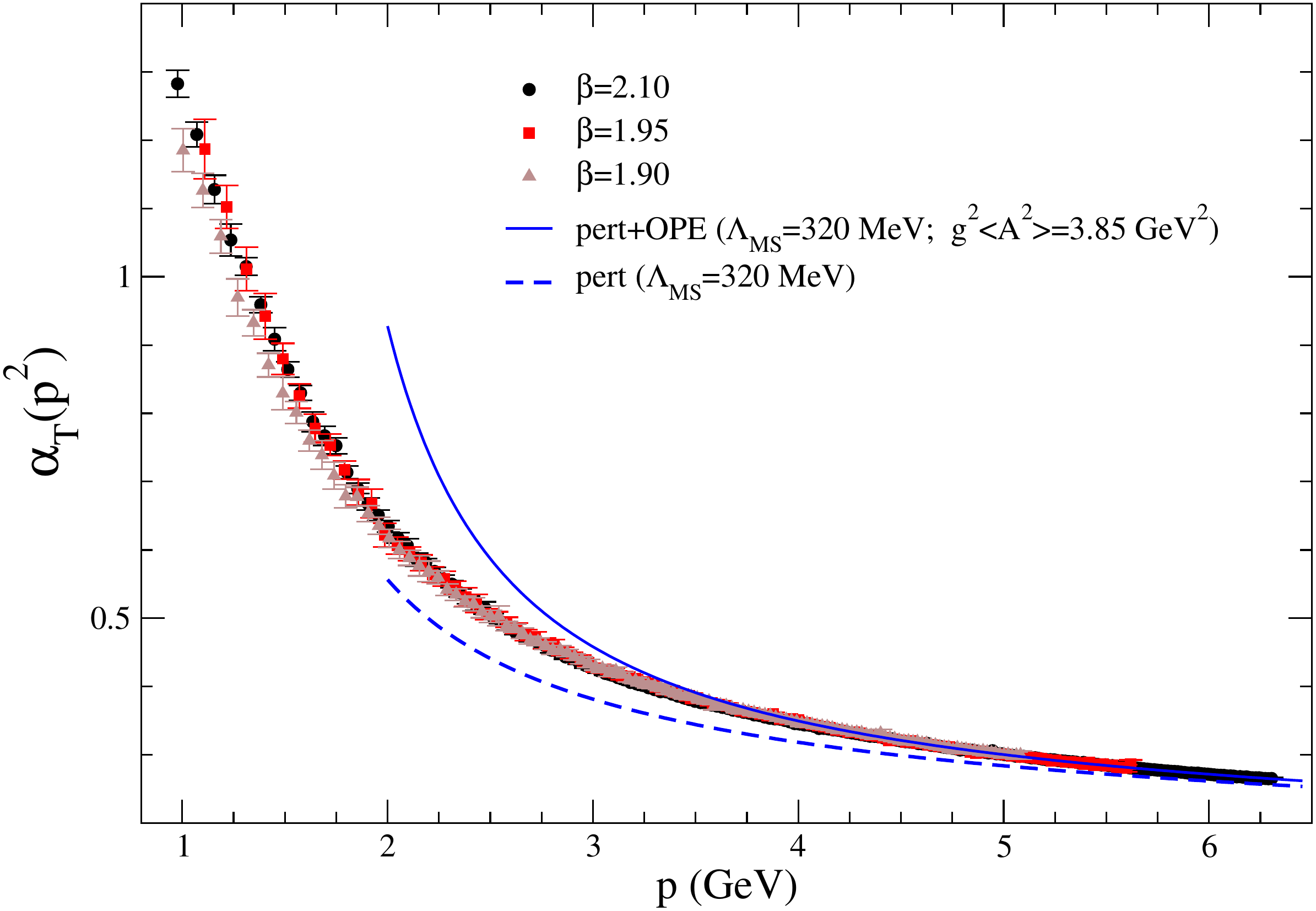}      
\end{center}
\caption{\small The Taylor coupling computed with the artefact-free lattice dressing functions and with \eq{eq:alphaTNP}  evaluated with  the best-fit parameters for $\Lambda_{\msbar}$ and $g^2\langle A^2 \rangle$ (solid blue line). For the sake of comparison, the four-loop perturbative prediction evaluated for the same $\Lambda_{\msbar}$ is also plotted with a dashed blue line. 
}
\label{fig:alpha}
\end{figure}

After applying the procedure to sweep the $O(4)$-invariant lattice artefacts away, we are left with the artefacts-free gluon and ghost dressing functions. One can then apply the definition, \eq{eq:alphaTdef}, to get an estimate of the Taylor coupling that is supposed not to be affected by artefacts at the order ${\cal O}(a^2p^2)$. In ref.~\cite{Blossier:2011tf}, we proceeded the other way around: we first computed the Taylor coupling from lattice data through \eq{eq:alphaTdef} and only then applied both the $H(4)$ extrapolation and 
$O(4)$-invariants artefacts removal to cure hypercubic artefacts at the desired order. Thus, 
re-obtaining the Taylor coupling, now after dropping the lattice artefacts away from the dressings, must be a strong consistency check for the artefacts treatment and for the whole analysis. As the artefacts-free dressing functions have been rescaled such that data for $\beta=1.90$ and $\beta=1.95$ result superimposed onto those for $\beta=2.10$ 
(see Fig.~\ref{fig:green}), we plug these rescaled dressings into \eq{eq:alphaTdef} and take the bare coupling, $g_0^2$, for $\beta=2.1$. The results are shown in Fig.~\ref{fig:alpha}, where a fit with \eq{eq:alphaTNP}, both $\Lambda_{\msbar}$ and $g^2 \langle A^2 \rangle$ taken as free parameters, is also plotted. One obtains for the fitted values (including jackknife errors): 
$\Lambda_{\msbar}=320(10)$ MeV and $g^2 \langle A^2 \rangle=3.9(3)$ GeV$^2$; strikingly in agreement with the results from the direct analysis of gluon and ghost dressings, in previous section, and with those from refs.~\cite{Blossier:2011tf,Blossier:2012ef}.

In the following, we will emphasize that obtaining compatible values of $\Lambda_{\msbar}$ and $g^2\langle  A^ 2 \rangle$ from both a fit of \eq{eq:alphaTNP} to the Taylor-coupling lattice data and fits of \eq{eq:GammaR1} to gluon and ghost dressings is neither trivial nor tautological. In the case of the renormalized Taylor coupling, once the artefacts been properly removed, not only the running with momenta is determined by the fitted parameters but also the size itself of the coupling at any momentum. 
To make it clear, \eq{eq:GammaR1} for the gluon ($\Gamma=G$) and ghost ($\Gamma=F$) dressings can be aso plugged into \eq{eq:alphaTdef} and one gets
\beq
\alpha_T(p^2) &=& \frac{g_0^2(a^{-1})}{4 \pi} \ z_{gh}^2\left(\mu^2,a^{-1}\right) 
z_{gl}\left(\mu^2,a^{-1}\right) \ 
\left( F_0^{\rm MOM}\left(\frac{p^2}{\mu^ 2},\alpha(\mu^ 2)\right) \right)^2 \ 
 G_0^{\rm MOM}\left(\frac{p^2}{\mu^ 2},\alpha(\mu^ 2)\right) \ 
 \left( \rule[0cm]{0cm}{0.7cm} 1 + \cdots \right) 
\nonumber \\
&=&
\frac{6}{4 \pi \beta} \ \frac{z_{gh}^2\left(\mu^2,a^{-1}\right) 
z_{gl}\left(\mu^2,a^{-1}\right)}{\alpha^{\rm pert}_T(\mu^2)} \ 
\alpha_T^{\rm pert}(p^2) \left( \rule[0cm]{0cm}{0.7cm} 1 + \cdots \right) \ ;
\label{eq:relMOM2}
\eeq
where $z_{gl}$ ($z_{gh}$) is the overall factor for the gluon (ghost) dressing function determined  by fitting the $O(4)$-invariant artefacts with \eq{eq:art} and reported in Tab.~\ref{tab:res}. The brackets with dots inside trivially correspond to the same bracket accounting for the gluon condensate OPE correction in \eq{eq:alphaTNP} and, in obtaining the second-line r.h.s., the latice bare coupling definition, $g_0^2=6/\beta$, and \eq{eq:relMOM} are also applied. As \eq{eq:GammaRL} provides with a precise description for the gluon and ghost lattice dressings with $\Lambda_{\msbar}$ from $\tau$ decays and 
the gluon condensate reported in Tab.~\ref{tab:res}, over $4.5 < p < 6.3$ GeV, \eq{eq:relMOM2}'s r.h.s. necessarily provides with the momentum behaviour within the same window and for the same parameters. However, \eq{eq:alphaTNP} will be only re-obtained if 
\beq\label{eq:relMOM3}
\alpha_T^{\rm pert}(\mu^2) \ \equiv \frac{6}{4 \pi \beta} \ z_{gh}^2\left(\mu^2,a^{-1}\right) 
z_{gl}\left(\mu^2,a^{-1}\right) \ ,
\eeq   
with $\alpha_T^{\rm pert}$ evaluated for the same $\Lambda_{\msbar}$. This is an additional strong constraint, involving the overall bare parameters $z(\mu^2,a^{-1})$, that can be only observed if lattice artefacts are properly treated and \eq{eq:GammaRL} indeed accounts for nonperturbative corrections such that $z_{gl}$ ($z_{gh}$) corresponds~\footnote{If so, \eq{eq:relMOM3} is nothing but the perturbative definition for the Taylor coupling.} to the  perturbative bare dressing function for the gluon (ghost). Then, the perturbative Taylor coupling can be computed from \eq{eq:relMOM3}'s r.h.s. at $\mu=10$ GeV, with the results reported in tab.~\ref{tab:res} and we thus obtain: $\alpha_T^{\rm pert}=0.216(2)$ for $\beta=2.10$, $\alpha_T^{\rm pert}=0.222(8)$ for $\beta=1.95$ and $\alpha_T^{\rm pert}=0.211(12)$ for $\beta=1.90$. These values can  be compared with $\alpha_T^{\rm pert}=0.214$, obtained in perturbation for $\Lambda_{\msbar}=311$ MeV,  or with $\alpha_T^{\rm pert}=0.216$ for $\Lambda_{\msbar}=320$ MeV. This comparison clearly provides with a pretty positive checking of the constraint given by \eq{eq:relMOM3}. 

\section{Conclusions}
\label{sec:conclu}

The gluon and ghost propagators, computed from lattice QCD simulations with two light and two heavy dynamical quark flavours, have been successfully described, for momenta above 4.5  GeV, with running formulae including four-loop perturbative corrections and a nonperturbative OPE power contribution led by the only dimension-two gluon condensate in Landau gauge, $g^2\langle A^2 \rangle$. The OPE formulae including only this leading nonperturbative correction fails to describe properly the lattice data for the propagators at momenta below 4.5 GeV, where next-to-leading corrections appear to be required. The contribution from the nonperturbative correction to the running is given by the Wilson coefficients for the local operator $A^2$ in the OPE
expansions of both gluon and ghost two-point functions. After defining the appropriate renormalization scheme for the two-point function and the local operator, the Wilson coefficient is also known at the four-loop order.  
As $\Lambda_{\msbar}$ for $N_f$=2+1+1 is well known from $\tau$ decays or can be consistently obtained from the world average value for the Strong coupling at $Z_0$ mass scale, the remaining lattice artefacts can be removed and the nonperturbative contribution isolated from data on the same footing. {\it This allows for a precise and positive test of the running due to the Wilson coefficients from data, the gluon condensate value being the same for both gluon and ghost two-point functions}. The main results of the paper appear thus sketched in Tab.~\ref{tab:comp}, where the universality for the condensate is checked, and in Fig.~\ref{fig:Wilson}, where we scrutinize the gluon and ghost propagators data to isolate the contribution from the Wilson coefficient to their running with momenta.

We finally tested that $\alpha_T$ directly derived from the bare ghost and gluon dressing functions is consistent with the fits on both the renormalized dressing functions; and also compatible $\Lambda_{\msbar}$ obtained from $\tau$ decays  (as shown in ref.~\cite{Blossier:2012ef}). This confirms in a non trivial way the validity of our estimates for the lattice artifacts and the nonperturbative correction, \eq{eq:GammaRL}. 

In summary, we found unequivocal deviations for ghost and gluon lattice propagators with respect to their four-loop perturbative prediction that have been consistently and accurately accommodated within the nonperturbative OPE approach. 

\appendix

\section{The matching criterion}

The criterion to determine the best matching for the artefacts-free rescaled dressing functions, simulated at different 
$\beta$ parameters, is based on the minimization of the area between polynomial fits of data. To this purpose, we 
fit the data with Legendre polynomials, 
\beq\label{eq:GPbeta}
\frac{z(a(\beta_0))}{z(a(\beta))} \ G(q^2,a^{-1}(\beta)) \ = \ 
\sum_{i=0}^n w_i(\beta) \ P_i\left(2 \frac{q-q_{\rm min}}{q_{\rm max}-q_{\rm min}} - 1\right)  \ = \ 
f_\beta(q) \ , 
\eeq
where $P_i(x)$ is the $i$-th order Legendre polynomial that, being defined within the interval $(-1,1)$, 
appears in \eq{eq:GPbeta} written in terms of an argument that makes it to range from $q_{\rm min}$ to $q_{\rm max}$, 
the lower and upper bounds for the fitting window.  The coefficient $w_i(\beta)$ corresponds to the weight for the 
$i$-th order polynomial. Therefore, we will have:
\beq\label{eq:area}
A(\beta_1,\beta_2) \ = \ \frac{\displaystyle \int_{q_{\rm min}}^{q_{\rm max}} \ dq \left( f_{\beta_1}(q) - f_{\beta_2}(q) \right)^2} 
{\displaystyle  \int_{q_{\rm min}}^{q_{\rm max}} \ dq \  \left( f_{\beta_1}(q)\right)^2}
\ = \ 
 \frac{\displaystyle \sum_{i=0}^n \frac{\left( w_i(\beta_1)-w_i(\beta_2)\right)^2}{i+\frac 1 2}}
{\displaystyle \sum_{i=0}^n \frac{ w_i(\beta_1)^2}{i+\frac 1 2}}
  \ .
\eeq
For our three lattice data sets, we will take $\beta_0=2.10$ and will fix $n+1=10$ (the number of terms for the Legendre polynomials), $q_{\rm min}=2$ GeV and $q_{\rm max}$ will be the minimum of the largest momenta for each two data sets being matched.
Then, we finally define 
\beq
A_p \ = \ A(2.10,1.95) + A(2.10,1.90)
\eeq
as the parameter to be minimized.

\section*{Acknowledgements} 

We thank the support of Spanish MICINN FPA2011-23781 and 
``Junta de Andalucia'' P07FQM02962 research projects, and  
the IN2P3 (CNRS-Lyon), IDRIS (CNRS-Orsay), TGCC (Bruyes-Le-Chatel), CINES (Montpellier) 
and apeNEXT (Rome) computing centers. Numerical calculations have also benefited from HPC resources of GENCI (Grant 052271) and CCIN2P3K. K. Petrov is part of P2IO Laboratory 
of Excellence.



\end{document}